\documentclass[twocolumn,twoside,slac]{revtex4}
\usepackage{graphicx}
\usepackage{fancyhdr}
\begin{document}

\title{Auger \& XtremWeb: Monte Carlo computation on a global computing platform}

% Repeat the \author .. \affiliation  etc. as needed
%
% \affiliation command applies to all authors since the last
% \affiliation command. The \affiliation command should follow the
% other information

\author{Oleg Lodygensky}
\affiliation{LAL, Paris South University, France}
\author{Gilles Fedak}
\affiliation{LRI, Paris South University, France}
\author{Vincent Neri}
\affiliation{LRI, Paris South University, France}
\author{Alain Cordier}
\affiliation{LAL, Paris South University, France}
\author{Franck Cappello}
\affiliation{LRI, Paris South University, France}

\begin{abstract}

In this paper, we present XtremWeb, a Global Computing platform used to generate monte carlos showers in Auger \footnote{Auger is an HEP experiment to study the highest energy cosmic rays at Mallargue-Mendoza, Argentina}.

XtremWeb main goal, as a Global Computing platform, is to compute distributed applications using idle time of widely interconnected machines. It is especially dedicated to -but not limited to- multi-parameters applications such as monte carlos computations; its security mechanisms ensuring not only hosts integrity but also results certification and its fault tolerant features, encouraged us to test it and, finally, to deploy it as to support our CPU needs to simulate showers.

We first introduce Auger computing needs and how Global Computing could help. We then detail XtremWeb architecture and goals. The fourth and last part presents the profits we have gained to choose this platform. We conclude on what could be done next.

\end{abstract}

\maketitle

\thispagestyle{fancy}

% body of paper here - Use proper section commands
% References should be done using the \cite, \ref, and \label commands
% Put \label in argument of \section for cross-referencing
%\section{\label{}}

%=Section================================================
\section{Introduction}
\label{sec:introduction}
%=Section================================================

The aim of the Pierre Auger Observatory is to detect showers produced by the interaction of cosmic rays of energy greater than 10E+19eV with the atmosphere. In order to determine the origin of these cosmic rays, their direction and energy must be measured with accuracy. Their nature (photons, protons or nuclei) must be known too.

These measurements are based on the properties of the secondary particles of the shower reaching the ground (number, position, energy, nature and mean arrival time) and on the study of the nitrogen fluorescence generated by these particles through the atmosphere. However statistical fluctuations in the development of an air-shower exist. For that reason, the data analysis needs a large number of simulated air-showers, with a good accuracy, to study these fluctuations. As duration of one simulation is about 10 hours, the needed simulation computing time is estimated at 106 hours for all the experiment.\\

The Aires\cite{AIRES} (Air-shower Extended Simulations) is one of the main program of simulation used by the Auger Collaboration; it is already used in Computing Center of the IN2P3 in Lyon. As computer sciences evolve, it appeared that we can now use new technology to break Computer Center barriers and gain computing resources distributed among the Internet.\\
These new possibilities are known as {\it Global Computing}.

%=Section================================================
\section{Global computing}
\label{sec:GC}
%=Section================================================

Global computing is an intensive computer science research field which aim is to distribute and share computing resource (CPU, disk space etc.). Among these proposals are different approaches which all try to solve problems on large scale computing. This last is a paradigm that addresses the problems of resource sharing between distributed systems in a dynamic, flexible, secure and non disturbing way, from within different organizations (universities, companies etc.) or even individuals.\\

The different studies around global computing are generalized as Grid computing, but can be divided into at least two groups, peer-to-peer computing (P2P) and Grid \cite{GRID0}.

``Grid'' main goal is to achieve flexible and secure large scale computing with high performances between so called {\it virtual organizations}, entities that accept a resource exchange policy, based on high control about who shares what, what is shared and what are the sharing conditions. The main toolkit, Globus \cite{GLOBUS}, is already used by several projects. Its strong security mechanism (GSI) is one of its main contributions. However, Globus lacks a mechanism for transparent fault tolerance that leads to use or implement a fault tolerance environment in top of Globus.

``P2P'' has quite the same goal, to achieve flexible and secure large scale computing with high performances, but in a more decentralized way. P2P system main features are resource volatility and the lack of security mechanism for resources, applications and their results. Some P2P deployments have already shown useful performances (SETI@Home for P2P computing with several Teraflops, Kazaa for P2P file sharing with one Terabits/s of service bandwidth...) and fault tolerance capabilities (the time between two connections or disconnections is lower than a minute).

%=Section================================================
\section{XtremWeb}
\label{sec:XtremWeb}
%=Section================================================

XtremWeb is a P2P project developed at University of Paris-Sud, France\cite{XTREMWEB}. It was originally designed to study execution models in the general framework of Global Computing and is now a full production platform too; it has been released for Linux, Windows and MacOS-X.

Distribute applications among set of resources (i.e. hosts) is a widely spread idea that leads different team to work on (SETI@Home, Folding@Home), and some projects especially focus on distribution of multi parameter applications, which need to be executed several times with different input/parameters, each computation being independent from each other. Among such projects is Nimrod \cite{Nimrod} which uses a static set of resources and which security relies on standard Unix level security. We see in following paragraphs that XtremWeb focuses on multi parameters applications too, but uses dynamic resources accordingly to their availability and implements its own strong security policy.

%=SubSection=============================================
\subsection{Design}
\label{sec:Design}
%=SubSection=============================================

XtremWeb implements three distinct entities, the {\it coordinator}, the {\it workers} and the {\it clients}, to create a so-called XtremWeb {\it network} (i.e. the Global Computing platform) using connectionless protocols only.\\
The coordinator masters the tasks management process (tasks scheduling and results storage) and is the only piece under the full control of the XtremWeb network administrator (i.e. the user who creates an XtremWeb network). Clients are software instances available for any user allowed to submit tasks to the XtremWeb network; it submits tasks to the coordinator, providing binaries and optional parameter files and permits the end user to retrieve his results. Finally, the workers are the software part spread among volunteer hosts to compute tasks. Everything is written in {\it Java} language for portability purposes.\\

XtremWeb protocols, as defined below, resolve firewall problems by using single side communications. Firewalls are usually configured asymmetrically allowing outgoing connections and blocking incoming ones. Workers and clients behind a firewall or even a gateway implementing {\it NAT} can then contact the coordinator and receive answers through the same opened canal. The coordinator, which is the only one in XtremWeb to open incoming ports, can receive connections since it uses the standard {\it Web} port (80) and firewalls are usually configured to let incoming connections to this port. If the firewall in front of the coordinator stops these connections, this will be the only one to be reconfigured so that the full system works.

\paragraph{The coordinator.} Tasks are managed following the {\it coordinator-worker} paradigm. One host (the {\it coordinator}) manages a bag of tasks provided by {\it clients}, coordinates their scheduling among a set of hosts (the {\it workers}) that are volunteers provided by institutional or private users and, as such, are not under the control of the coordinator and are very volatile in essence. Following this concept, each action is initiated by workers only. This behavior is commonly known as {\it pull} model and clearly implies independence of all components.\\
Scheduling is in {\it FIFO (first in, first out)} mode. XtremWeb can schedule native and {\it Java} applications, so there's a match done on CPU type, OS version and whether Java is enabled in workers. Java applications are distributed as {\it jar} files, whereas native ones are  binary.\\
Tasks are scheduled to workers on their specific demand only since they may appear (connect to coordinator) and disappear (disconnect from coordinator) with no predictable pattern (a worker is then said {\it connected} as long as it periodically contacts the coordinator). Any scheduled task is expected to be computed by a worker and have its results sent back to the coordinator; on failure, the task is re-scheduled to another worker.

\paragraph{The clients.} Clients are distributed to authorized users only to make them able to submit tasks to the coordinator as transactions. Before submitting any task, the client contacts the coordinator to fetch any previous submitted ones. This ensures that when the client restarts from a fault or any other reason, it does not resubmit previously submitted tasks. Results are managed according to the user needs. They can be discarded immediately after fetch or kept by the coordinator until the end of the session. So on client failure, it is the responsibility of the client programmer to fetch relevant results. An {\it API} is implemented to provide such secure implementations.

\paragraph{The workers.} Workers are distributed entity to volunteer institutional or individual PCs, which aim to use CPU accordingly to a local user customizable policy ({\it available scheduling time, CPU usage conditions...}) to compute tasks provided by the coordinator. A worker requests task to compute accordingly to its own local policy; it downloads task software and all expected objects (input file, arguments...), stores them on reliable media and starts computing the provided tasks. Computation goes on locally until it ends or dies for any reason, including due to host utilization policy rules. As computation is started, the worker periodically signals the coordinator, so that last knows the computation goes on well. If unable to connect (coordinator or network shut down), the worker still continues computation and will signal the coordinator as soon as possible.\\
After any failure (network or host shut down, or local usage policy rule) the worker retrieves its current state; it restarts the interrupted task, if any, from the beginning as there is no {\it checkpoint}\cite{Pruitt} implemented in XtremWeb and signal the coordinator the task it is computing. A worker may then be asked to stop computing, if it task has been rescheduled.\\
When a task is completed, the worker sends results back to the coordinator and ask for another one.\\

%=SubSection=============================================
\subsection{Fault tolerance}
\label{sec:Fault}
%=SubSection=============================================

Several fault tolerance mechanisms are used in XtremWeb to handle clients, workers and coordinator failures. The main purpose of these mechanisms is to enable the system to restart properly after any failure (worker, client and coordinator). It is not currently intended to provide minimum service interruption using techniques like redundancy, but is planned as future work.\\

The coordinator manages its tasks using transactions and stores them in reliable media (disk) so that the full system integrity is preserved even if the coordinator shuts down for any reason (crash or system management). At starting time, the coordinator reads the information stored on reliable media to set up its proper state; it then retrieves tasks (scheduled and awaiting ones). The client submits tasks and the worker fetches tasks using transactions. This ensures a consistent state when the coordinator restarts from fault while the client and the worker have not failed.\\
Worker faults are detected thanks to the {\it alive} signal and their tasks may then be rescheduled on another available worker. A worker can impromptingly be told to stop its current task if it has been disconnected for too long (i.e. if it has not signaled the coordinator in time) to avoid redundant task and, more, result overwriting.\\

XtremWeb then achieve to prevent fault tolerance transparently for user, using resilient components fetching their context before restarting.

%=SubSection=============================================
\subsection{Security}
\label{sec:Security}
%=SubSection=============================================

XtremWeb has the responsibility to ensure user authentication, hosts (workers) integrity, application and results protection and user execution logging.

Security mainly relies on three completer mechanisms : a list of
authorized users as ACLs managed by the coordinator, authentication of
the coordinator by  workers and clients and self-protection of
workers by the use of sand boxing system utility.

 All tasks are submitted to the workers through the coordinator credential
 and contain a descriptor with the actual user identity so that workers
 and coordinator can take appropriate corrective action (user
 revocation), in case of security problem. 
 Therefore it is a first concern to ensure that workers and clients
 connect to the appropriate coordinator. To do so, all communications to
 the coordinator are performed within a SSL tunnel. To open the
 connection a challenge is run using the public key (certificate) of the
 coordinator, previously inserted within the worker or client software.
 We suppose that the download and installation  of the worker/client is
 a safe stage in the process. As the coordinator certificate is the only
 one contained in the list of trusted  keys (the list itself cannot be updated) , this prevents the ``Man
 in the Middle'' attack, because the attacker would have to own the
 private key of the coordinator to let the SSL session to be established.  This mechanism prevents malicious
 participants to be able to intercept and read any connection, and to
 connect to the coordinator; it also prevents workers/clients to connect to a
 wrong XtremWeb coordinator.
 On the contrary client and worker are not authenticated by the
 coordinator beside the use of the couple login/password.  This point requires
 the maintenance of a database which could be evicted by integrating a
 certificate/signature system like PKI or GSI in Globus.

 XtremWeb workers protect their host by implementing {\it
 sand boxing}\cite{SANDBOX2,SANDBOX3,SANDBOX1} to secure binary
 application executions, providing rights to do some actions and denying
 some others since binary applications have access to the full hosting
 system by nature. Workers are configured to run any task of that type
 inside a sand box which is fully customizable, from memory usage to file
 system operations. Java applications, on their side, are always
 executed inside a virtual machine which includes
 security\cite{JAVASECURITY}; XtremWeb uses this functionality in two
 levels, one for the worker itself and a more restrictive one for the
 downloaded Java byte code. Java virtual machines, as sand boxes, have a
 performance cost\cite{SANDBOX-PERF}.\\

 Finally, XtremWeb must ensure application, parameters and results
 integrity. Even if all these objects (binary, parameters and results)
 are safely transferred, they are still vulnerable inside the worker host
 itself as they are stored onto disk. The major difficulty then comes
 from two points. The first is that anybody (or at least the host owner)
 have the full access of hosting machine, even XtremWeb worker temporary
 files (binary, parameters and results). The second point is on XtremWeb
 open source policy, which makes protection coding ineffective since
 anybody can download the source code. The XtremWeb team is currently
 working on this problem and will present results on paper to come.

 Host and connection security is achieved in XtremWeb that can then be
 deployed being sure volunteer hosts integrity is guaranteed. There's
 still applications and their results to be certified and, as previously
 said, these subjects are under work by the XtremWeb team.

\begin{figure*}[t]
\vspace{0.5cm}
\centering
\includegraphics[width=90mm]{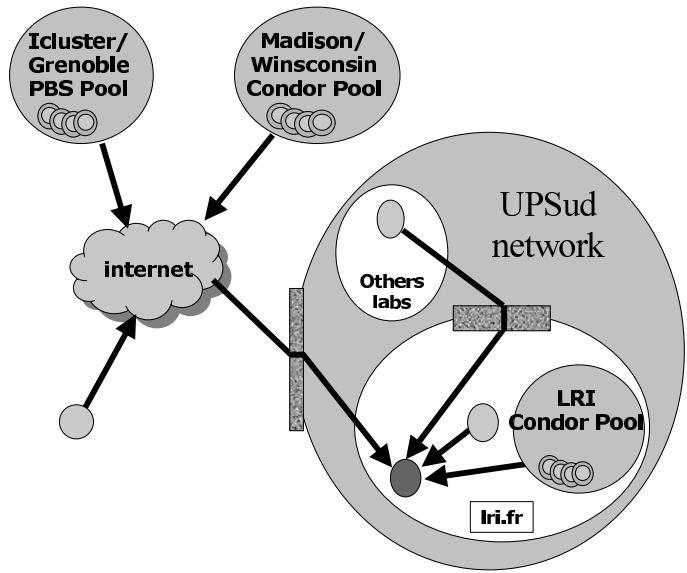}
\caption{XtremWeb experimentation configuration.} \label{config}
\end{figure*}

%=Subsection================================================
\subsection{User experiences and feedback}
\label{sec:Feedback}
%=Subsection================================================

XtremWeb is already used by several projects as listed below:

\begin{itemize}
 \item CGP2P  ACI GRID (academic research on Desktop Grid systems), France 
 \item Industry research project (Airbus + Alcatel Space), France
 \item IFP (French Petroleum Institute), France
 \item EADS (Airplane + Ariane rocket manufacturer), France
 \item University of Geneva, (research on Desktop Grid systems), Switzerland 
 \item University of Wisconsin Madison, Condor+XW, USA
 \item University of Guadeloupe + Paster Institute (Research on tuberculosis), France 
 \item Mathematics lab University of Paris South (PDE solver research) , France 
 \item University of Lille (control language for Desktop Grid systems), France
 \item ENS Lyon: research on large scale storage, France 	
\end{itemize}

%=Section================================================
\section{Auger simulations distributed with XtremWeb}
\label{sec:Feedback}
%=Section================================================

Prior to any proposal to introduce XtremWeb into Auger simulated showers management, we decided to make some experimentations. We introduced 1024 identical Aires tasks into the XtremWeb network to determine how the full system reacts. Each task is computed in about 18 minutes on the reference host, a mono-processor PIII 733Mhz. These 1024 tasks would take 307 hours on that host.

Figure~\ref{config} shows the experimentation platform we used to test XtremWeb to distribute simulation computations over three different sites: two sites in France, one at the LRI and the other at Grenoble, and one site in Wisconsin, USA.

The coordinator run on a dedicated machine at LRI. Workers run on different sites, managed by batch systems to make the deployment just easier; we deployed then our XtremWeb workers as tasks locally managed by provided batch systems. Grenoble site uses PBS~\cite{feitelson97theory} whereas Wisconsin and LRI sites use Condor\cite{webCondor}. As these two batch systems use different resource allocation policies, no prediction can be made about how and when our workers are scheduled. Figure \ref{config} also shows some workers not included in any cluster. This is to make clear that XtremWeb workers don't need to be managed by any cluster and may be run on any personal computer. Our experiments used workers on clusters only to keep management as easiest as possible.\\

We have made five experiments : WISC-97 (97 processors at Wisconsin), WL-113 (113 processors at Wisconsin and LRI), G-146 (146 processors at Grenoble), WLG-270 (270 processors at Wisconsin, LRI and Grenoble) and WLG-451 (451 processors at Wisconsin, LRI and Grenoble).

\begin{figure*}[t]
\centering
\includegraphics[width=90mm]{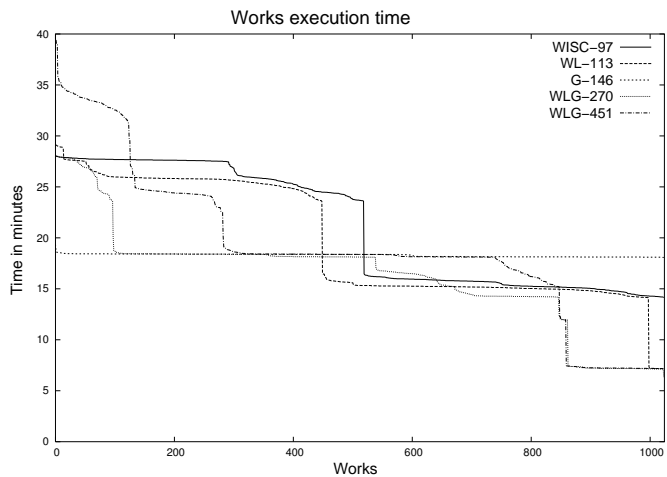}
\caption{Work execution time.} \label{cpu}
\end{figure*}

Figure \ref{cpu} shows works execution time for each task, decreasingly sorted. We note the remarkable stability of the system at Grenoble (G-146 curve) where all hosts are identical (CPU, memory...) whereas WISC-97 clearly shows two levels corresponding on the two available host types (PIII 533Mhz and 900Mhz).
That last does not show the stability found in G-146. This is because Condor may simultaneously allocate the same resource (i.e. the same CPU) to different tasks, depending on the local resource management policy and, in another hand, because the network bandwidth is not the same between Wisconsin and LRI, and between Grenoble and LRI.

\begin{figure*}[t]
\centering
\includegraphics[width=90mm]{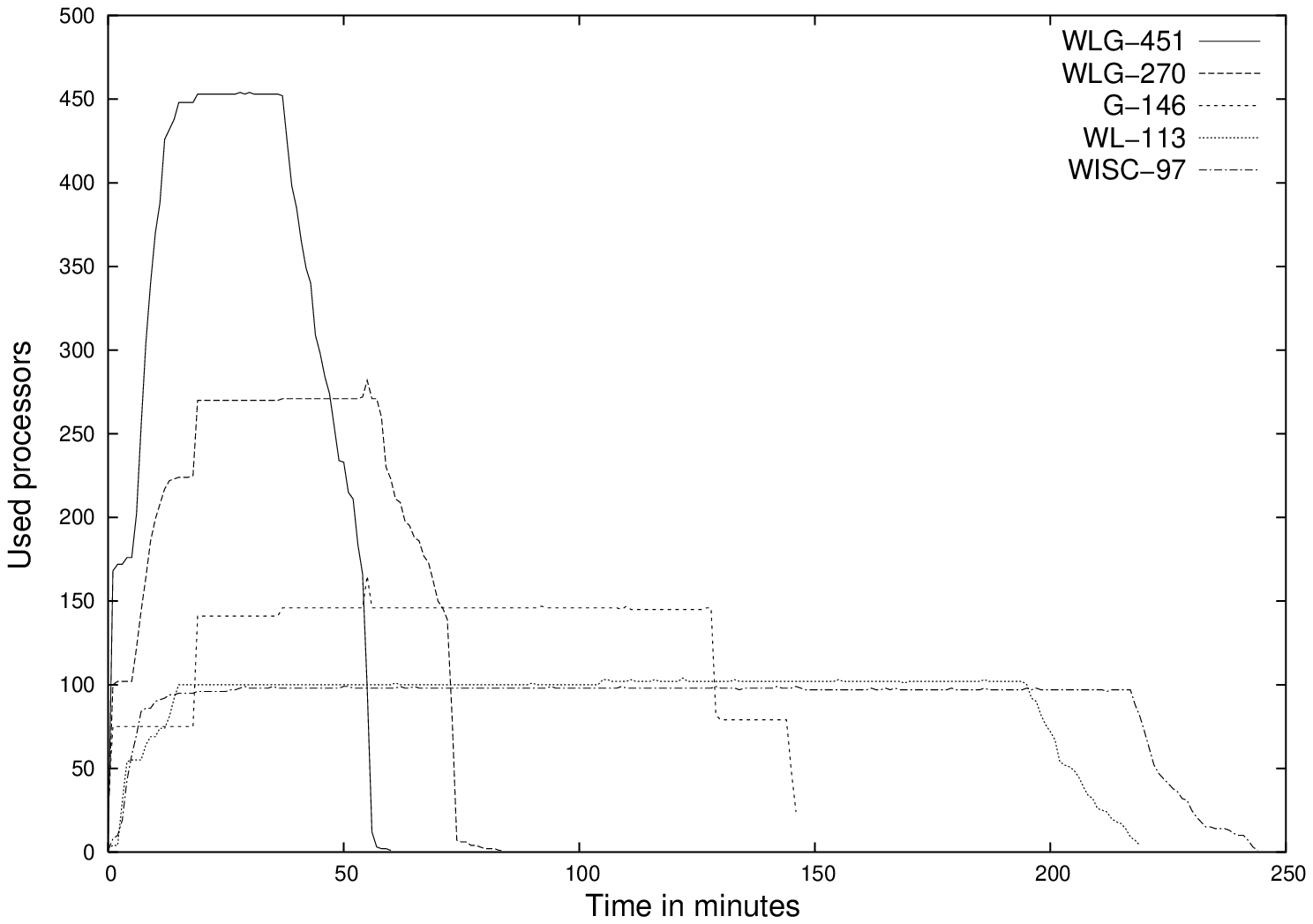}
\caption{Processors utilization.} \label{procs}
\end{figure*}

Figure \ref{procs} shows resource utilization for each experimentation. All curves show three different steps. The first is the power slope as CPU are allocated by clusters; then we reach the efficient power utilization which finally decreases as experimentation ends.

%=Section================================================
\section{Conclusion and future work}
\label{sec:Conclusion}
%=Section================================================

\begin{figure*}[t]
\vspace{0.5cm}
\centering
\includegraphics[width=110mm]{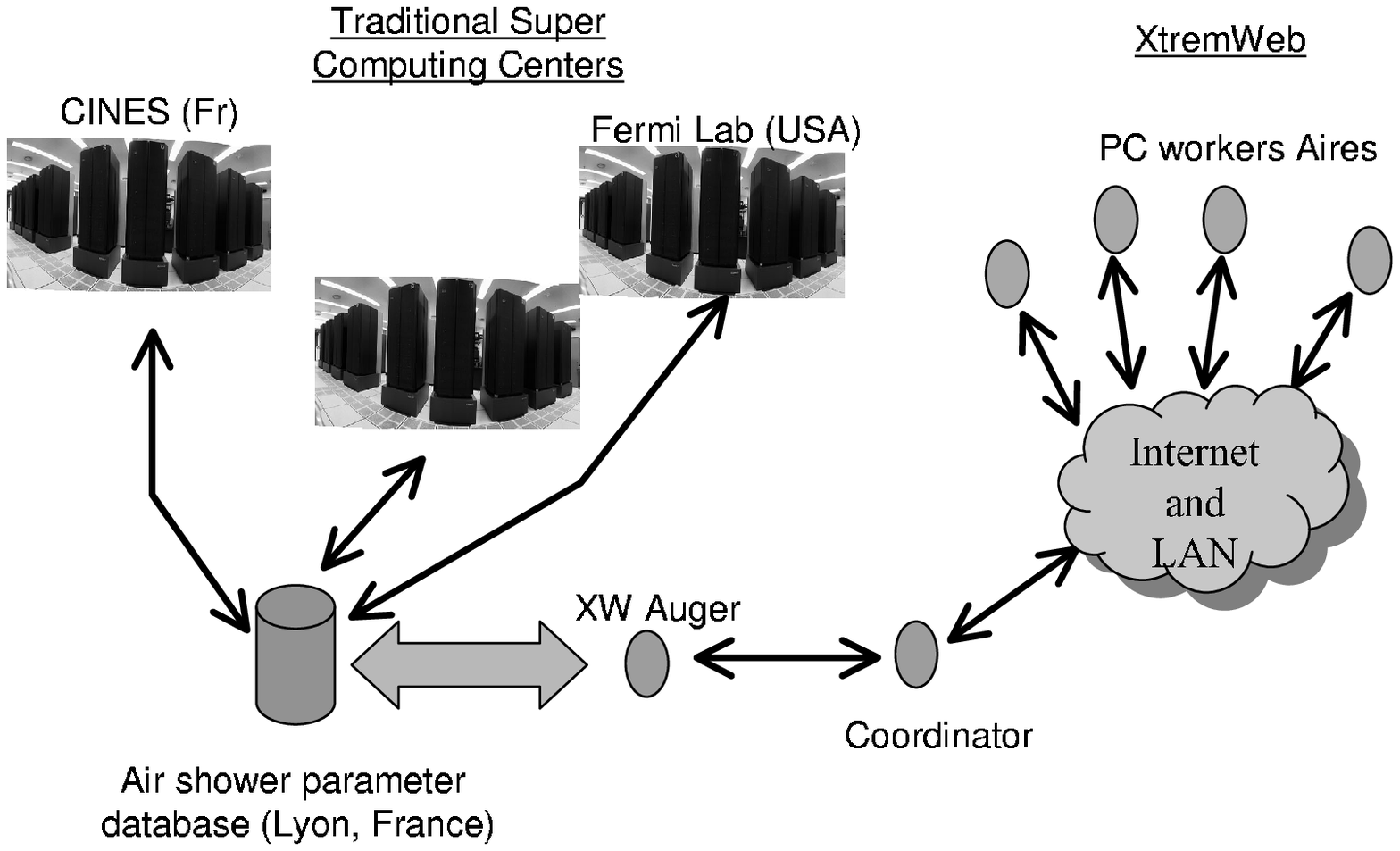}
\caption{XtremWeb configuration for Auger.} \label{auger}
\end{figure*}

Figure \ref{auger} shows Auger configuration for Monte Carlos computations as we expect it to be in near future, as tests are fully convincing. We can see that Auger will use two different ways to compute its simulated showers; the first one uses ``standard'' CPU power provided by traditional computing centers whereas the other implements an XtremWeb network. Both are connected to Auger database ({\it Auger DB}) at Lyon, France to get informations about showers and to generate and store generated showers.\\

Parts will be deployed as follow:

\begin{itemize}
 \item a daemon ({\it Xw Auger}) as XtremWeb client; it scans Auger DB for new showers to be generated and submits them to the coordinator. On the other hand, its periodically connects to the coordinator to retrieve results, if any;
 \item the coordinator is then not connected by itself to Auger DB; it definitely knows nothing specific to Auger and only sees a client ({\it Xw Auger}) as it would for any XtremWeb client as previously defined;
 \item the workers are widely distributed among volunteer collaboration hosts. They download {\it Aires} binary, a shower simulator, compute the shower and send results back to the coordinator. Again, this XtremWeb component has just nothing specific to Auger.
\end{itemize}

The first phase will accept worker connections from collaboration hosts only since the XtremWeb results certification is not available yet.\\

We have presented a Global Computing platform, XtremWeb, which usage experiment has presented an easy and convenient way to answer our CPU needs for monte carlos computations. This platform achieves to resolve most of the problems encountered by any Grid system such as scalability, fault tolerance and security.

As our tests have completed, we believe that this platform responds to the expectations of Auger users who would easily deploy workers among PCs of the collaboration so that an effective XtremWeb network could grow and propose a non-negligible CPU power.

%%\bibliography{../Bibliography/Mybib.bib}

\end{document}